\documentclass[aps,twocolumn,superscriptaddress]{revtex4-1}
\usepackage{amsmath,amssymb}
\usepackage{graphics,graphicx}
\usepackage{dcolumn,bm}
\usepackage{psfrag}
\usepackage{color}
\usepackage{multirow}
\usepackage{braket}
\newcommand{\cu}
{\affiliation{Department of Physics, University of Calcutta,
92 Acharya Prafulla Chandra Road, Kolkata 700009, India.}}

\begin{document}
\title{Persistent quantum walks: dynamic phases and diverging timescales}

\author{Suchetana Mukhopadhyay}
\cu
\author{Parongama Sen}%
\cu

\begin{abstract}
A discrete time quantum walk is considered in which the step lengths are chosen to be either
$1$ or  $2$  with the additional feature that the walker is persistent with a 
probability $p$. This implies that with  probability $p$, the walker repeats 
the step length taken in the previous step and is otherwise  antipersistent. 
We estimate the probability $P(x,t)$ that the walker is at $x$ at time $t$ and the first two moments.  
Asymptotically, 
$\langle x^2 \rangle = t^\nu$ for all $p$.
For the extreme limits $p=0$ and $1$, the walk is known to show ballistic behaviour, i.e., $\nu = 2$.
As $p$ is varied from zero to 1, the system is found in four  different phases characterised by  the  value of $\nu$: $\nu =2$ at $p=0$, $1 \leq \nu \leq  3/2$ for $0 < p < p_c$,  $\nu = 3/2$ for $ p_c < p <1$ and $\nu = 2$ again at $p=1$. $p_c$ is found to be very close to $1/3$ numerically. 
Close to $p=0,1$, the scaling behaviour shows a crossover in time. Associated with this crossover, two diverging  timescales     varying as $1/p$ and $1/(1-p)$ close to $p=0$ and $p=1$ respectively are detected. 
Using a different scheme in  which the antipersistence behaviour is suppressed, one gets $\nu= 3/2$  for the entire region $0 < p< 1$. 
Further, a measure of the entropy of entanglement is studied for both the schemes.
\end{abstract}

\maketitle

\section{Introduction}
\label{sec:intro}
Discrete time quantum walks (DTQW's), first introduced by Aharonov $et\:al$.\cite{aharo}, are random walks where a coin degree of freedom is introduced which 
determines the translation of the walker. Quantum interference in such walks leads to the position $x$ of the walker scaling as $\langle x^2\rangle \propto t^2$ with time $t$, indicating a quadratically faster spread than the classical random walk.

The introduction of randomness or disorder in quantum walks has been demonstrated to modify its scaling behavior
significantly. Disorder can be incorporated in various ways, and several studies in recent years have focused on the modifications they impart and how they may turn out to be useful \cite{kendon}. 
Disorder introduced through interaction with the external environment or the presence of broken links in position space tends to 
slow down the quantum walk  and leads to localization \cite{keating,yin}. Such localization effects were first studied by Anderson \cite{anderson} in the context of electron localization in a disordered lattice. Anderson localization type behavior has been observed experimentally by introducing static (positional) disorder in a quantum walk on a homogeneous lattice \cite{schreiber}. Static disorder can be contrasted with dynamic disorder or decoherence \cite{yin} that transforms the quantum walk to the classical equivalent (diffusive scaling; $\langle\,x^2\,\rangle - {\langle\,x\,\rangle}^2 \propto t$).

Dynamic disorder is usually introduced through the operations controlling the evolution of the quantum walk, such as by using decoherent coins \cite{brun2,brundeco}. It is also possible to incorporate dynamic disorder by relaxing the standard assumption of a constant displacement at each time step and allowing longer steps to be chosen randomly, as in \cite{psen,psen1} where the scaling $\langle\,x^2\,\rangle\propto\,t^{\frac{3}{2}}$ was found. 

In the present work we introduce the concept of persistence in the quantum walk. In a classical random walk, persistence implies that the walker continues in the direction taken in the previous step, making it non-Markovian. In the quantum walk, in order to introduce the idea of persistence, we allow the walker to take different step lengths at each step 
and remember the step length chosen in the preceding step
with a certain probability. 
 This provides a simple way to study the effect of short term memory in the long-ranged walk. 
The probability distribution of the position of the walker and its first and second moments are evaluated and the results compared with the classical walk and the quantum walk without disorder. In addition, we evaluate the entropy of entanglement.

Previously, a few studies have been made where memory has been incorporated in different ways in a quantum walk \cite{mcgett,rohde,limemory,molfetta,mol2}. Both short term and long term memory have been considered but in a  very  different manner. In particular, in the case of the so-called `elephant quantum walk' \cite{molfetta} where the walker has infinite memory as well as time dependent step lengths, the variance was found to scale as    $t^{3}$. 
To the best of our knowledge, the persistence in quantum walks, the way it has been incorporated in the present work, has not been considered before. 

In the next section we introduce the quantum walk and the exact way 
the concept of persistence has been used. In Section III, the results have been presented. Section IV includes   a summary of the results and a detailed discussion on the implications and the insight developed through the study. 

\section{The Persistent Quantum Walk}
\label{sec:TSDTQW}

In the simple DTQW in one dimension, the walker can occupy discrete, equispaced sites $x$ on the real line and takes a step at unit time intervals. In addition to the position, the walker is assigned a second degree of freedom, by means of a coin state (either left ($\ket{L}$) or right ($\ket{R}$).   The state of the walker is described by the following two-component vector expressing probability amplitudes for the coin states:\\
\begin{equation}
\label{10}
\ket{\psi(x,t)} = \langle{x}|{\psi(t)}\rangle = \begin{bmatrix} a(x,t)\\ b(x,t)\end{bmatrix}.\end{equation}\\
The occupation probability of the site $x$ at time step $t$ is given by $P(x,t) = |\langle{x}|{\psi(t)}\rangle|^{2} = |a(x,t)|^{2}+|b(x,t)|^{2}$ with  the total probability is equal to $1$ at each time step. 
A step in the quantum walk consists of a rotation in the coin space followed by a translation. A standard choice for this rotation operator is the Hadamard coin $H$, given by

\begin{equation}
H = \frac{1}{\sqrt{2}}\begin{bmatrix}1 & 1 \\1 & -1 \end{bmatrix}.
\end{equation}

Instead of defining the step length to be a constant $l$, we allow it to be chosen from a binary distribution; $l(t) = \{1,2\}$ at any given time $t$. The conditional translation operator at a time $t$ is then written as
\begin{equation}
\begin{split}T(t) = \ket{R}\bra{R}\otimes\displaystyle\sum_{x}\ket{x+l(t)}\bra{x+l(t)}\\ + \ket{L}\bra{L}\otimes\displaystyle\sum_{x}\ket{x-l(t)}\bra{x-l(t)}\end{split}
\end{equation}
Allowing for a non-unique step length in this way enables one  to study the phenomenon of persistence by considering the tendency to adopt the step length used in the previous time step. This choice is made in two ways, outlined under two different schemes, I and II. 
In each scheme, at $t = 0$, the step length $l(0) = 1$ or $l(0) = 2$ is chosen with equal probability. In Scheme I, at any later time $t \neq 0$, the walker either chooses the same step length as in the previous time step (persistent) and otherwise necessarily chooses the other step length (anti-persistent).
In Scheme II, the walker is  persistent with a probability $p$, but this time with probability $(1-p)$, either of the step lengths  $l = 1$ and $l = 2$ are  chosen,  with probability $q$ and $(1-q)$ respectively.

For both schemes the walker is initialised with  $a(x,0) = b(x,0) = \frac{1}{\sqrt{2}}\delta_{x,0}$ 
which gives an asymmetric probability distribution profile in the absence of disorder.
The walk is evolved for 20000 time steps for all parameter values. We investigate how the occupation probability, moments, and entanglement depend on the 
parameter(s) used in the two schemes. All results are averaged over 4000 configurations.

\section{Results}

\subsection{Scheme I}

In this subsection we present the results for the first scheme considered. 
The walker here   chooses the step length taken in the previous step 
with probability $p$ and with probability $(1-p)$ it chooses strictly the 
other length.  The latter case thus corresponds to an antiperstent choice.

\subsubsection{Probability distribution}

When $p = 0$, steps of length $l = 1$ and $2$ are taken  alternately, with a possible sequence of steps given by $1,2,1,2,1...$ etc. The walk clearly has periodicity 2, and there is no randomness in the choice of steps. This particular walk has already been studied in \cite{psen} and was found to have the same scaling behavior as the ordinary quantum walk. The probability distribution resembles an overlap of distributions obtained for the ordinary walks with $l = 1$ and $2$ \cite{psen}.

\begin{figure}[h!]
 \includegraphics[width = 0.45\linewidth]{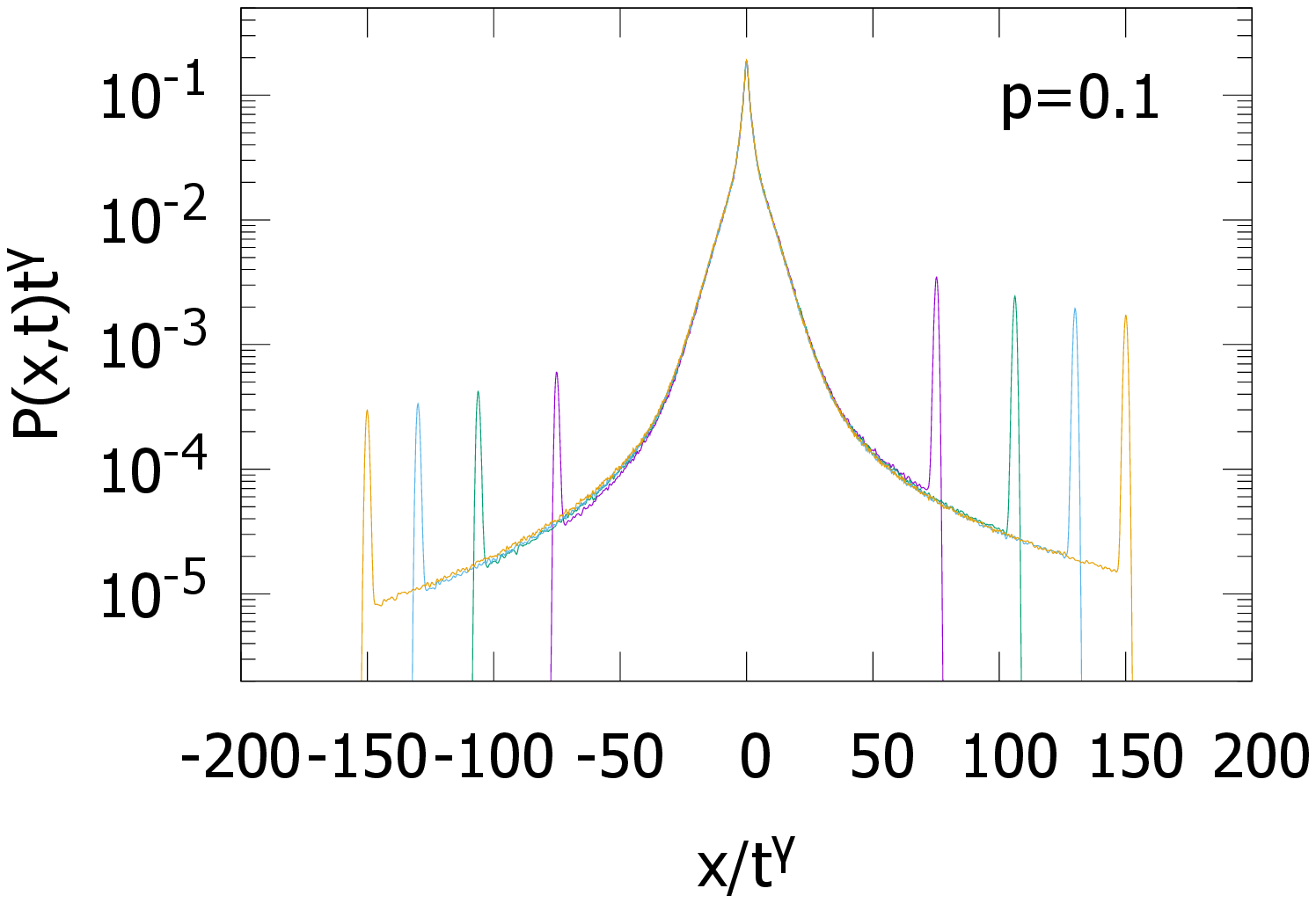}
\includegraphics[width = 0.45\linewidth]{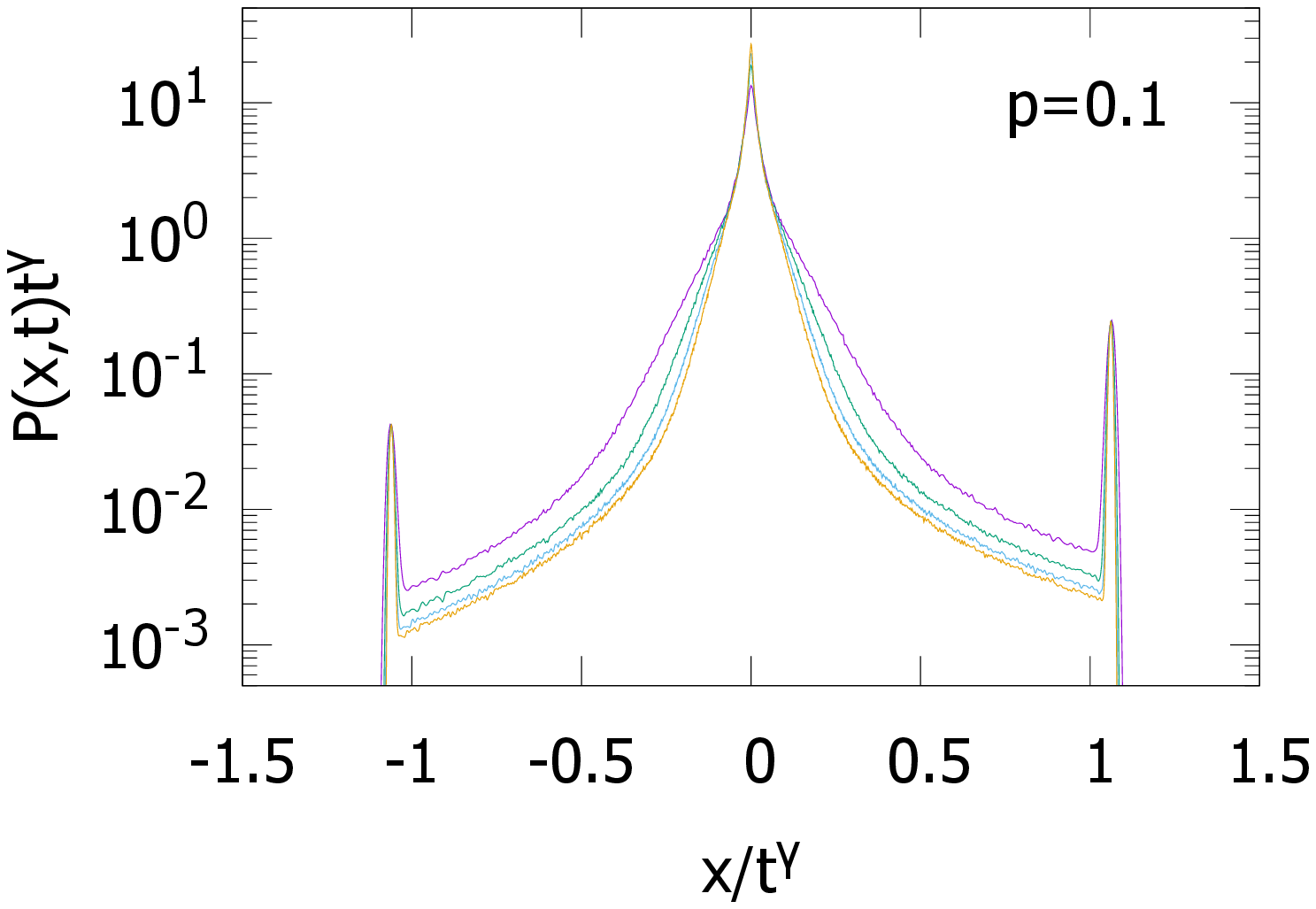}\\

\includegraphics[width = 0.45\linewidth]{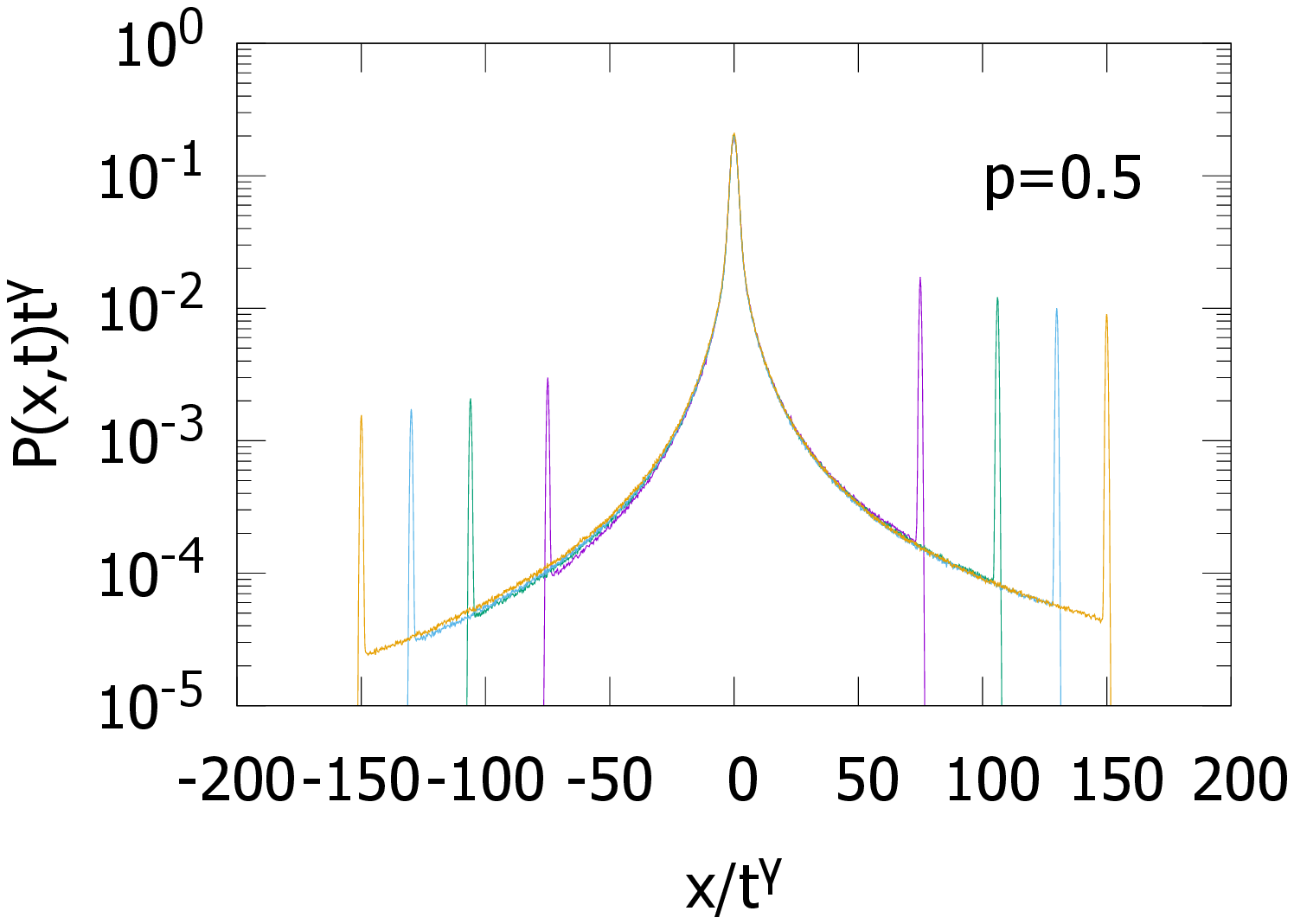}
\includegraphics[width = 0.45\linewidth]{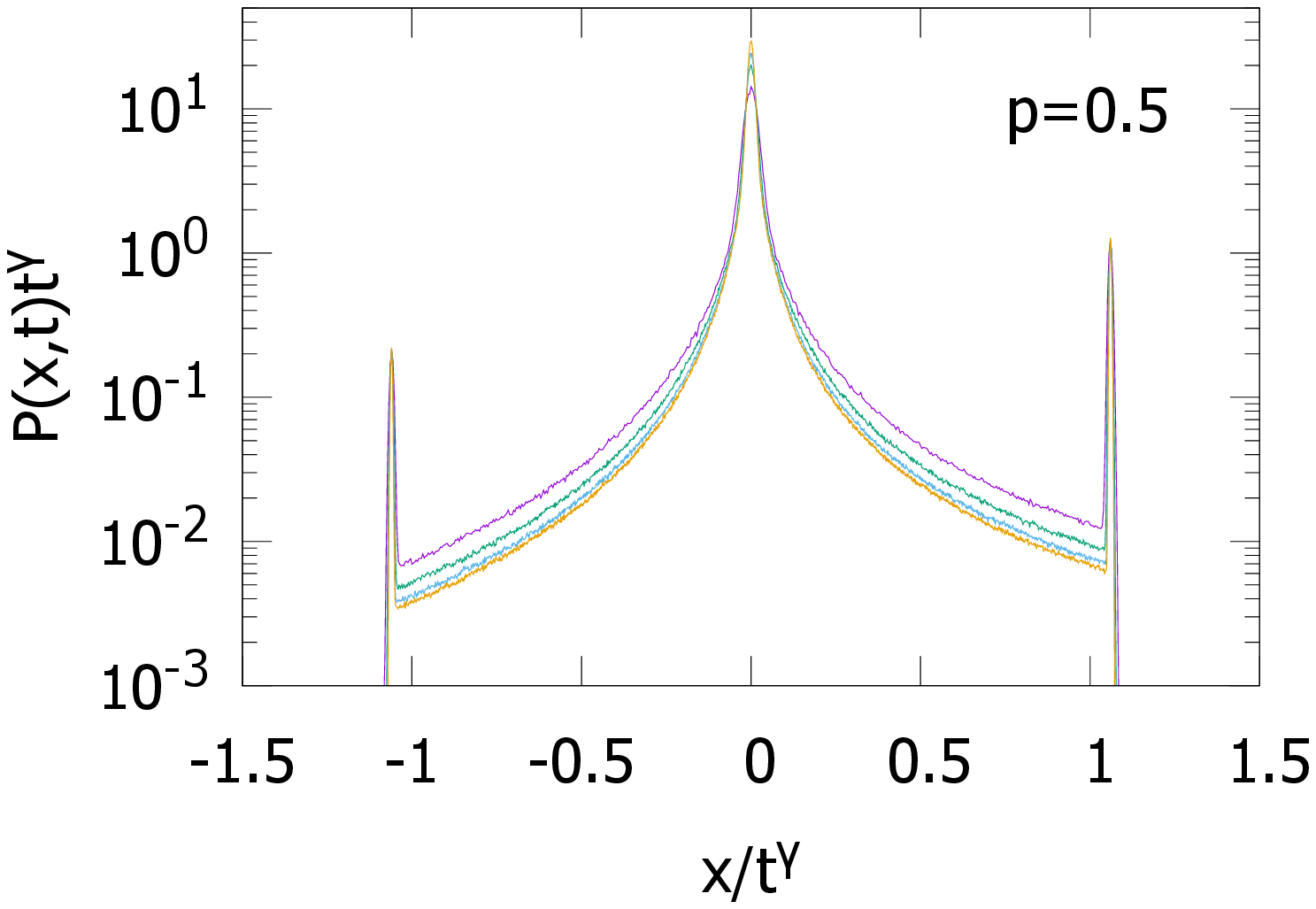}\\

\includegraphics[width = 0.45\linewidth]{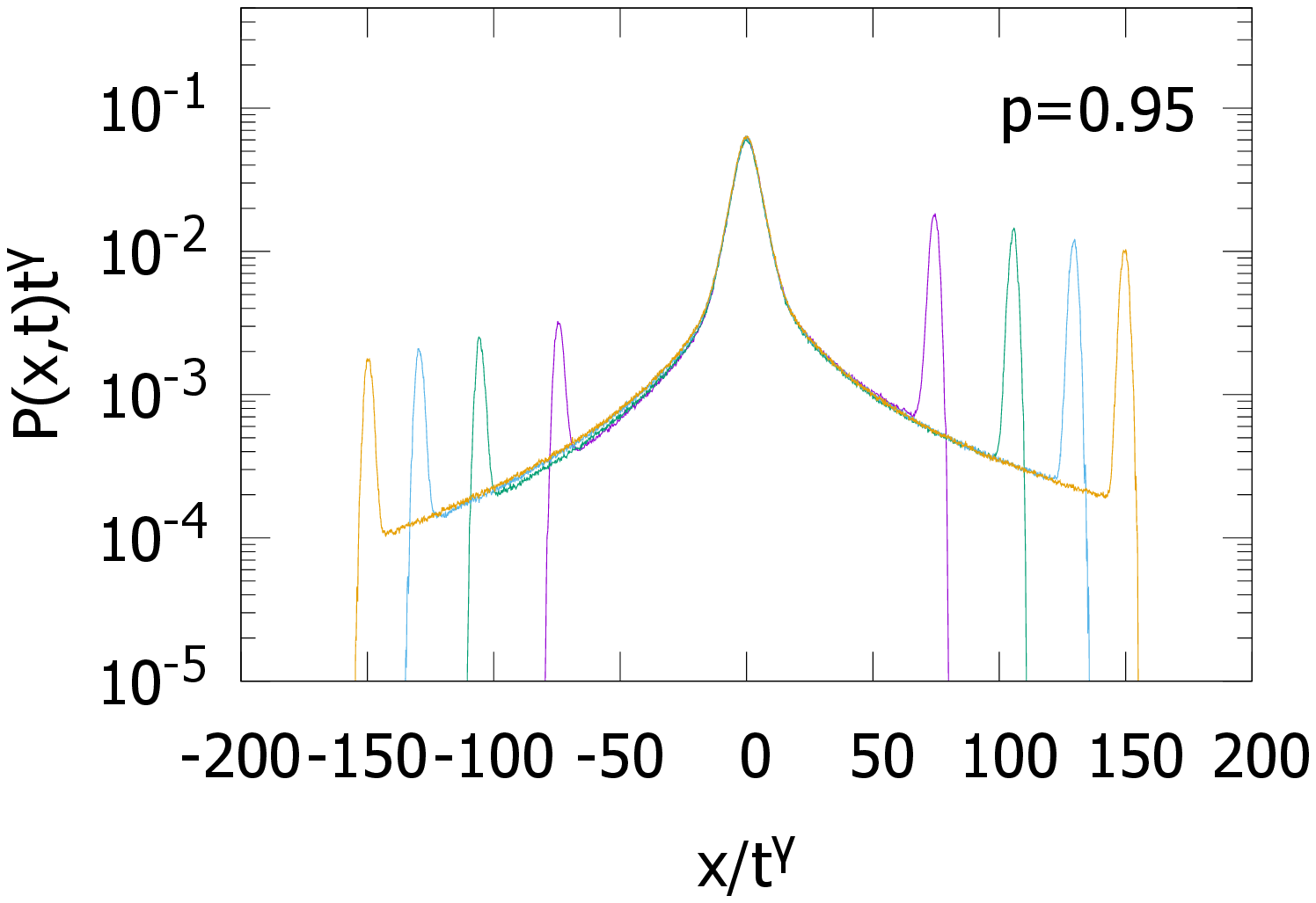}
\includegraphics[width = 0.45\linewidth]{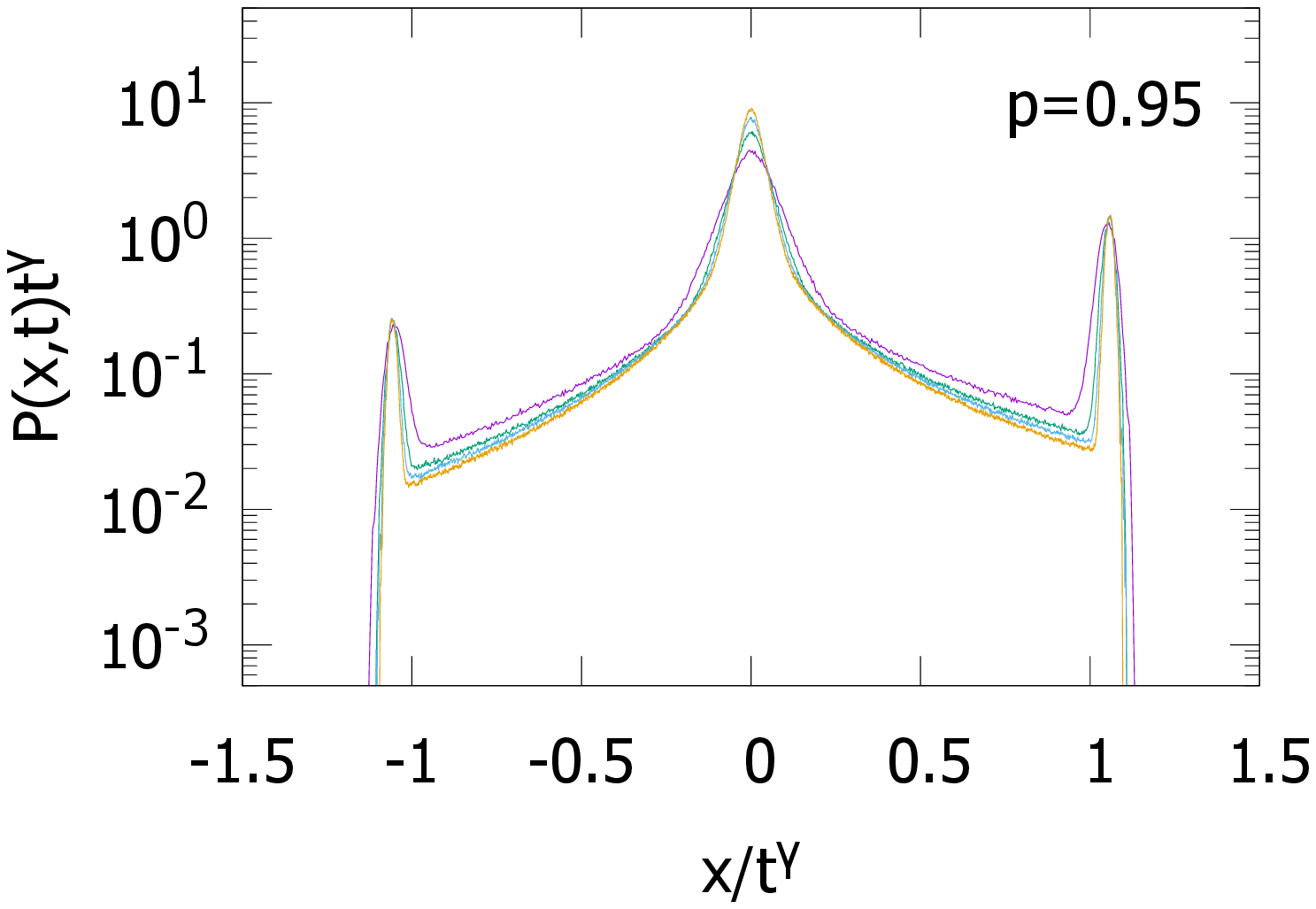}

\caption{Scheme I : Data collapse of rescaled $P(x,t)$  using $\gamma = 0.5$ (left column), $\gamma = 1.0$ (right column). $\gamma = 1$ collapse is less sharp.}
\label{fig1:dist-scheme1}
\end{figure}

On the other hand, when $p = 1$, the walk becomes deterministic with a  
unique step length and thus  identical to the usual quantum walk. 
The result for the point $p = 0.5$ can also be easily guessed. Here essentially the step lengths 1 and 2 are being chosen with 
equal probability at each step. Hence it belongs to the class of models studied in \cite{psen}. 

As $p$ is increased even slightly from zero, the distribution $P(x,t)$ 
 exhibits a  peak centered at the origin in addition to two ballistic peaks.  Such a central peak  
is not present in  the ordinary quantum walk, neither in the  binary  walk without randomness. However, 
in presence of disorder and decoherence, such peaks indeed appear,
when localisation of the quantum walker takes place.

The ballistic peaks are  signatures of the quantum nature of the walk, and are in general asymmetric in height, reflecting the asymmetry of the pure quantum  walk. As $p$ is increased, the ballistic  peaks are seen to increase in height, 
while the central peak goes down.  
This can be interpreted as an increase in delocalisation with increasing probability of the walker to be persistent, as the walker approaches the standard case of a constant step length. Even for $p$ very close to one, however, the central peak does not disappear. Exactly at  $p = 1$, the familiar distribution of the quantum walk is recovered, as expected.  

Plotting $t^{\gamma}P(x,t)$ against the scaled variable $x/t^{\gamma}$, we observe data collapse for four different time steps, with $\gamma =  0.5$ for the central peak and $\gamma = 1.0$ for the ballistic peaks. 
We conclude that $P(x,t)$ exhibits two distinct scaling behaviors: the centrally peaked part scales as $x \propto \sqrt{t}$, similar to the classical walk, while the ballistic peaks scale like the ordinary quantum walk, $x \propto t$.
These results are true for the entire region $0 < p < 1$ as shown in Fig. \ref{fig1:dist-scheme1}. 

\subsubsection{Scaling of the moments}

 We next present the results for  the first two  moments, 
$\langle x\rangle$ and 
$\langle x^2\rangle$ of the probability distribution. 

\begin{figure}
\centering \includegraphics[scale = 0.45]{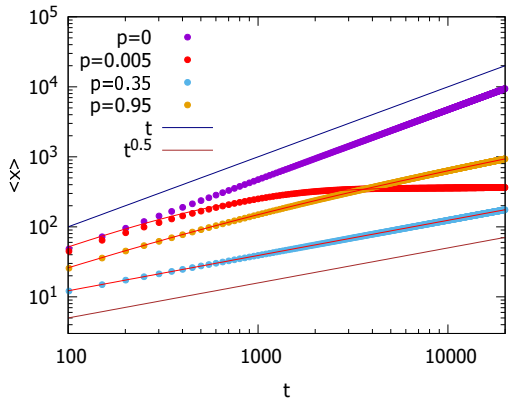}
\caption{Scheme I: The first moments for four $p$ values show significant change in behaviour as $p$ is varied.}
\label{fig2:fmom-scheme1}
\end{figure}

In the limiting cases of $p = 0$ and $p = 1$, the walk reduces to an ordinary quantum walk such that $\langle x\rangle$ $\propto t$ 
and 
  $\langle x^2  \rangle \propto t^2$.
Interestingly, when $p$ deviates from zero or one by even the smallest amount, we note that the asymptotic variations of the moments are significantly changed. 
The first moments are plotted in Fig.~\ref{fig2:fmom-scheme1} for a few $p$ values, showing 
that the asymptotic  exponent decreases to very small values for $p = 0^+$ and 
 increases up to a value close to 1/2 at larger $p$ values. 

\begin{figure}
\centering \includegraphics[scale = 0.45]{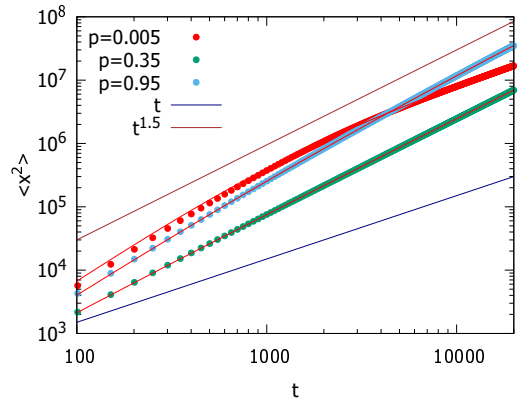}
\caption{Scheme I: The second moment for three $p$ values. The continuous lines are best fit curves obtained 
using Eq. \ref{crsvr} with fitting parameters $\alpha^\prime=1.379, 0.81, 1.81$ and $\beta^\prime= 0.0011, 0.382, 
0.066$ for $p = 0.005, 0.35$ and $0.95$ respectively.}
\label{fig3:mom-scheme1}
\end{figure}

We plot the second moments in Fig.~\ref{fig3:mom-scheme1} for a few values of $p$. 
Usually it is the variance which is used to characterise the walk. However, 
we note here 
that  asymptotically,  $\langle x^2\rangle$ is  larger than ${\langle x\rangle}^2$ by at least  two orders of magnitude. 
Hence it suffices to consider how $\langle x^2\rangle$ behaves 
instead of the variance $\langle x^2\rangle - {\langle x  \rangle}^2$. 
In the following, we study the behaviour of the second moment in detail. 

\begin{figure}
\centering \includegraphics[scale = 0.6]{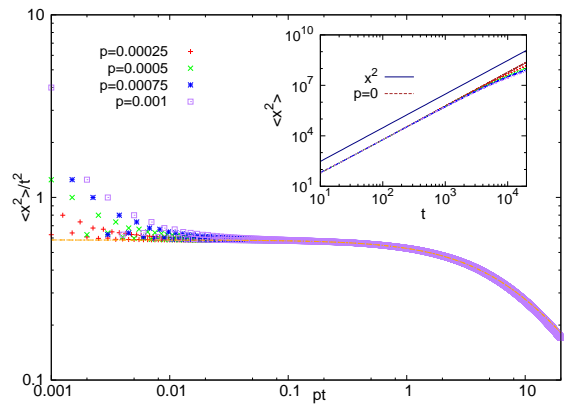}
\caption{Scheme I: 
Data collapse of the second 
 moment of the distribution $P(x,t)$
obtained on plotting $\langle x^2\rangle$/$t^2$ against the scaling variable $pt$. The solid line is a best fit line drawn using   Eq. \ref{scale0}. Inset shows the unscaled data $\langle x^2\rangle$ against time $t$. The data for $p=0$ and a curve with  quadratic variation  are shown for comparison.}

\label{fig4:scaling-scheme1p0}
\end{figure}

We note two things from Fig. \ref{fig3:mom-scheme1}; 
first, for $p$ values close to zero 0 and 1, there is a distinct change  in the behaviour of the moments with time;  initially it has a fast growth  but becomes slower later. Secondly, the asymptotic variation is significantly dependent on $p$; for a value of $p$ very close to zero, the exponent  is close to 1. This is a drastic change from the value 2 when $p$ is exactly zero.


We probe the $p \to 0^+$ region in more detail as we have the most significant change in the asymptotic exponent value here. 
Here, $\langle x^2 \rangle$ 
plotted in the inset of Fig. \ref{fig4:scaling-scheme1p0},
clearly shows a variation compatible with $t^2$ for a long time before deviating to a slower variation.
The deviation occurs at larger values of time as $p$ approaches zero. 
Plotting $\langle x^2\rangle$/$t^2$ against the scaling variable $pt$ for several values of $p$ very close to zero, we obtain a very good collapse (Fig. \ref{fig4:scaling-scheme1p0}) from which we claim $\langle x^2\rangle\propto\ t^2 f(pt)$. It is clear from   Fig. \ref{fig4:scaling-scheme1p0} that $f(z)$ fairly a  constant  for $z < 1$. For $z \geq 1$, one can 
fit $f(z) $ to the form 
\begin{equation}
f(z) = 1/(\alpha+\beta z^\mu)
\label{scale0}
\end{equation} 
to a great degree of accuracy with $\alpha \approx 1.7, \beta \approx 0.2$ and  $\mu \approx 1$. 
From this it can be  concluded  that a crossover occurs at $pt\approx 1$  and there is a diverging time scale varying inversely with $p$.  On the other hand the asymptotic variation is $ \langle x^2\rangle \propto t$. Thus the crossover  time marks the transition to the asymptotic behaviour.

\begin{figure}[h!]
\centering \includegraphics[scale = 0.5]{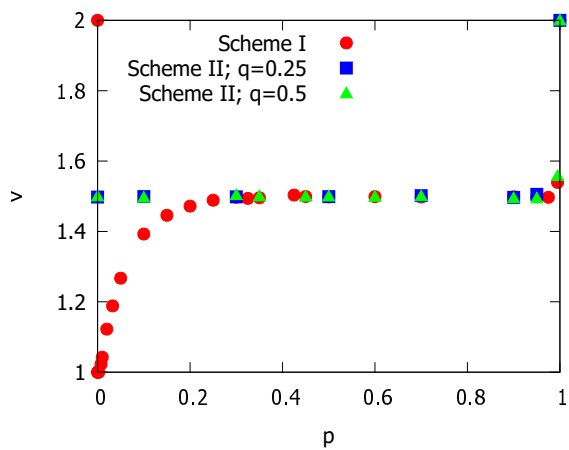}
\caption{Variation of the effective scaling exponent $\nu$ with $p$ for Scheme I and Scheme II. }
\label{fig5:nu}
\end{figure}

As $p$ is made larger, the crossover occurs at smaller times and the exponent is extracted from fitting the second moment directly to the empirical form valid for later times:
\begin{equation}
\label{crsvr}
\langle x^2\rangle = t^2/(\alpha^{\prime}+\beta^{\prime}\,t^{2-\nu}),
\end{equation}
such that asymptotically, $\langle x^2\rangle \propto t^{\nu}$. The best fit curves using the above form   
 plotted for the data shown  in Fig. \ref{fig3:mom-scheme1} show excellent agreement.
In Fig. \ref{fig5:nu}, the asymptotic exponents $\nu$ is plotted as a function of $p$. One notes that the exponent continuously varies from $1$ to $3/2$ in the region $0 < p < p_c$ where $p_c$ is approximately $0.33$.

A similar crossover is noted close to $p =1$, where also the exponent shows a 
jump from the value 2 to $\sim 1/2$. 
Here the scaling variable is $(1-p)t$ such that the associated timescale 
diverges as $\frac{1}{1-p}$.

We note therefore that  the walk shows a superdiffusive behaviour but with a nonuniversal exponent for small $p$ where the antipersistent effect is strong. 
For larger values of $p$, the walker behaves as that with random  step lengths and persistence is apparently merely the tool that  provides the stochasticity. Since it was found in \cite{psen}  that the slightest randomness alters the exponent to 3/2, it is not surprising to see that for large $p \neq 1$, one gets the same exponent.

\subsubsection{Entanglement entropy}

\begin{figure}[h!]
\includegraphics[width = 0.45\linewidth]{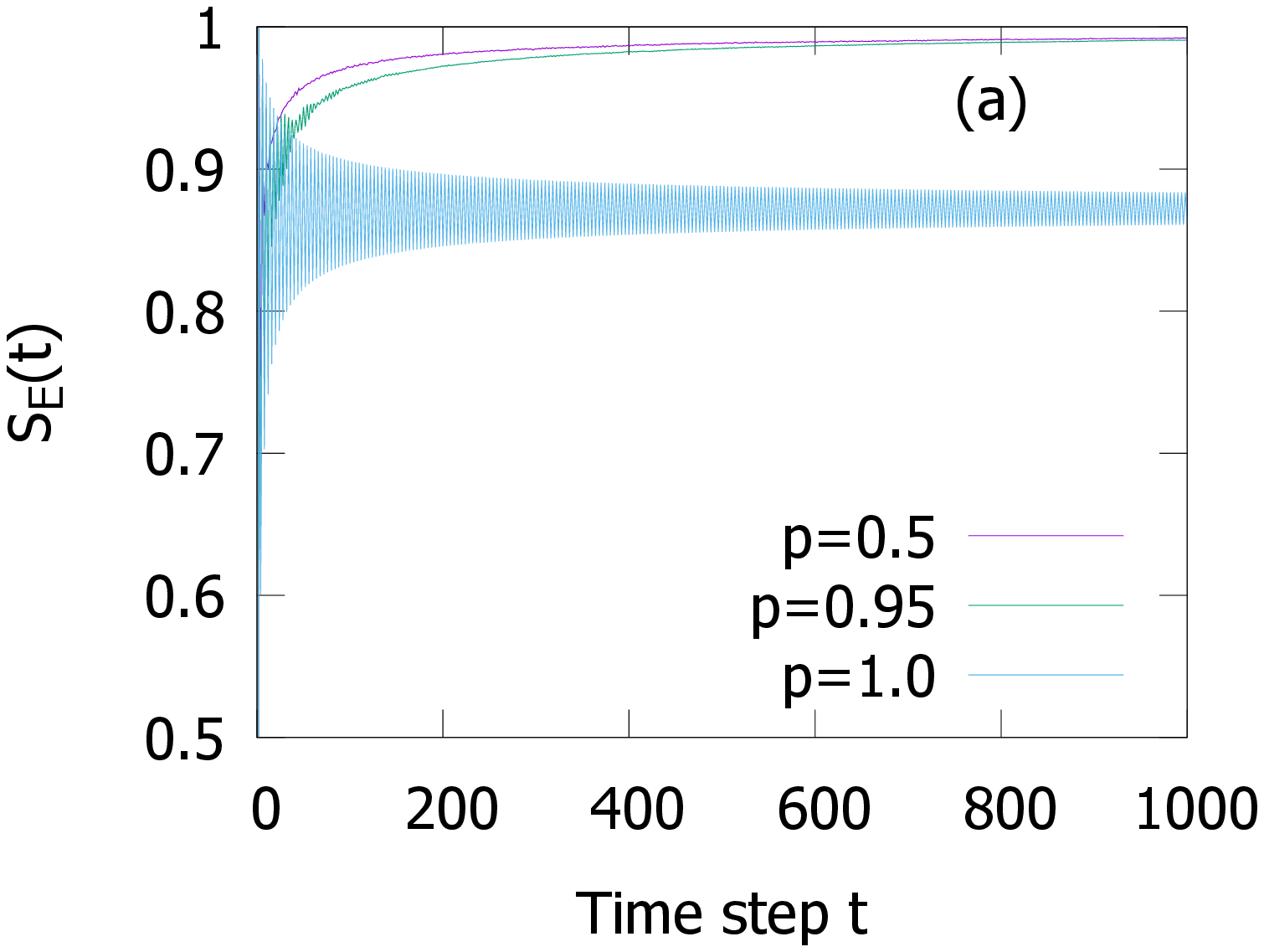}
\includegraphics[width = 0.45\linewidth]{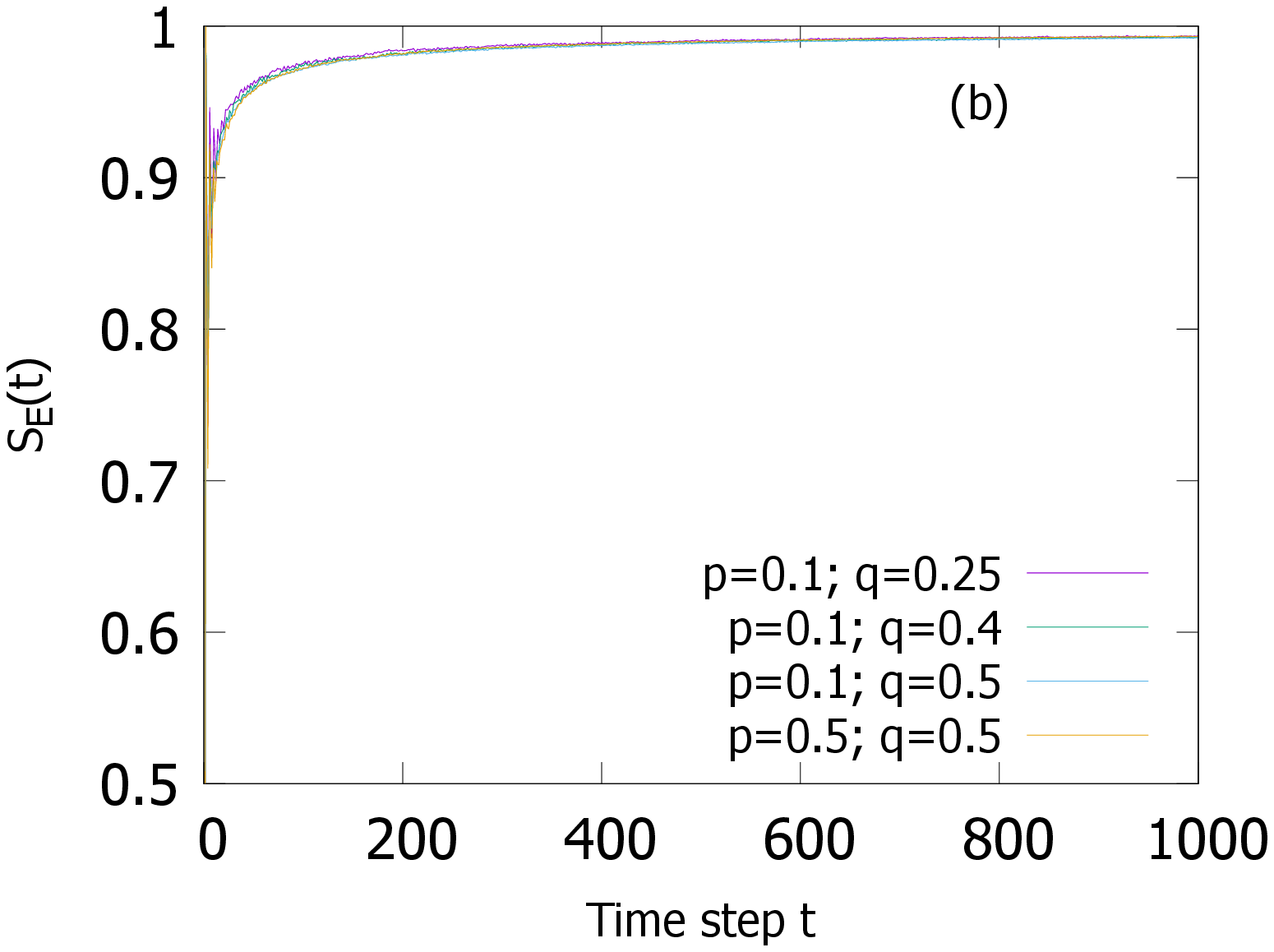}
\caption{Entanglement entropy plotted against time for Scheme I (a) and Scheme II (b).}
\label{fig6:entang}
\end{figure}

In any quantum walk, the evolution operator generates entanglement between the position and coin degrees of freedom. This entanglement can be quantified using the von Neumann entropy $S_E(t)$, also known as the entropy of entanglement. This can be evaluated from the reduced density operator which is represented by the following matrix \cite{abal} \begin{equation}\rho_c(t) = \begin{bmatrix}A(t) & B(t)\\ B(t) & C(t)\end{bmatrix}\end{equation} where we have $A(t)\equiv\sum_x |{a(x,t)}|^2$ ; $B(t)\equiv\sum_x |{a(x,t)}||{b(x,t)}|$ ; $C(t)\equiv\sum_x |{b(x,t)}|^2.$
The entropy of entanglement $S_E(t)$ is then calculated as
\begin{equation}\begin{aligned}S_E(t)&= -Tr(\rho_c\,\text{log}_2\,\rho_c)\\&= -(v_1\,\text{log}_2\,v_1\,+\,v_2\,\text{log}_2\,v_2).\end{aligned}\end{equation} where $v_1$ and $v_2$ are the real, positive eigenvalues of the matrix $\rho_c(t)$. We numerically evaluate $S_E(t)$ for the quantum walk with the localised initial condition $a(0,0) = \frac{1}{\sqrt{2}}$, $b(0,0) = \frac{1}{\sqrt{2}}$, taking $0\,\text{log}_2\,0 = 0$. For a constant step length, when $p = 1$, $S_E(t)$ is found to asymptotically converge to $\approx\,0.872...$ for our chosen initial condition, as has been previously reported \cite{carneiro, abal, abalerr}. \\
For $p = 0$, we obtain $S_E(t)\approx 0.85...$. As the walk deviates from either of the two extremes, the value of $S_E(t)$ increases drastically, rapidly converging to a large value very close to unity for any $p$. Although the parameter $p$ does not significantly influence the limiting value (at least not up to three decimal places), the rate of convergence is faster for smaller $p$ values. Fig. \ref{fig6:entang} shows the behavior of the entanglement for a few values of $p$.

\subsection{Scheme II}

This scheme represents a variation of the first where the walker is either persistent with probability $p$ or, with probability $(1-p)$ can choose step length $l = 1$ or $2$ with probability $q$ and $(1 - q)$ respectively. 
Obviously, for $p=1$ it is always persistent and one recovers an ordinary quantum walk.
For $p=0$, it takes step lengths 1 and 2 randomly unless $q =0$ or 1.
In fact, if $q=0$ or 1, the walk becomes of unique step length eventually, (independent of $p$); if $q=1$, that length is 1 and 2 for $q=0$.

 Effectively, the total persistence probability $p^\prime$ of the walker in this scheme is either $p + q(1-p)$ or $p + (1-q)(1-p)$ at a given time step and it is antipersistent with probability $1 - p^\prime$. When $q = 0.5$, the walker is thus  persistent with a probability $p^{\prime} = p/2 + 1/2$ independent of $q$, and the results should correspond to those obtained for a persistence probability  $p^{\prime}$ in Scheme I. Since $p^\prime > p_c$, one can
expect that the results for Scheme II will be identical to Scheme I with the second moment scaling as $t^{3/2}$ asymptotically for all $p \geq 0$.
Even when $q \neq 0.5$,  the walk is persistent with probability $p/2 + 1/2$ on an average and antipersistent otherwise.
Thus one can expect that it will  again be equivalent to  Scheme I (with persistence probability $p^\prime$), when ensemble average is taken if the fluctuation is negligible. In the following we discuss the results which confirms the above picture.


\begin{figure}[h!]
\includegraphics[width = 0.45\linewidth]{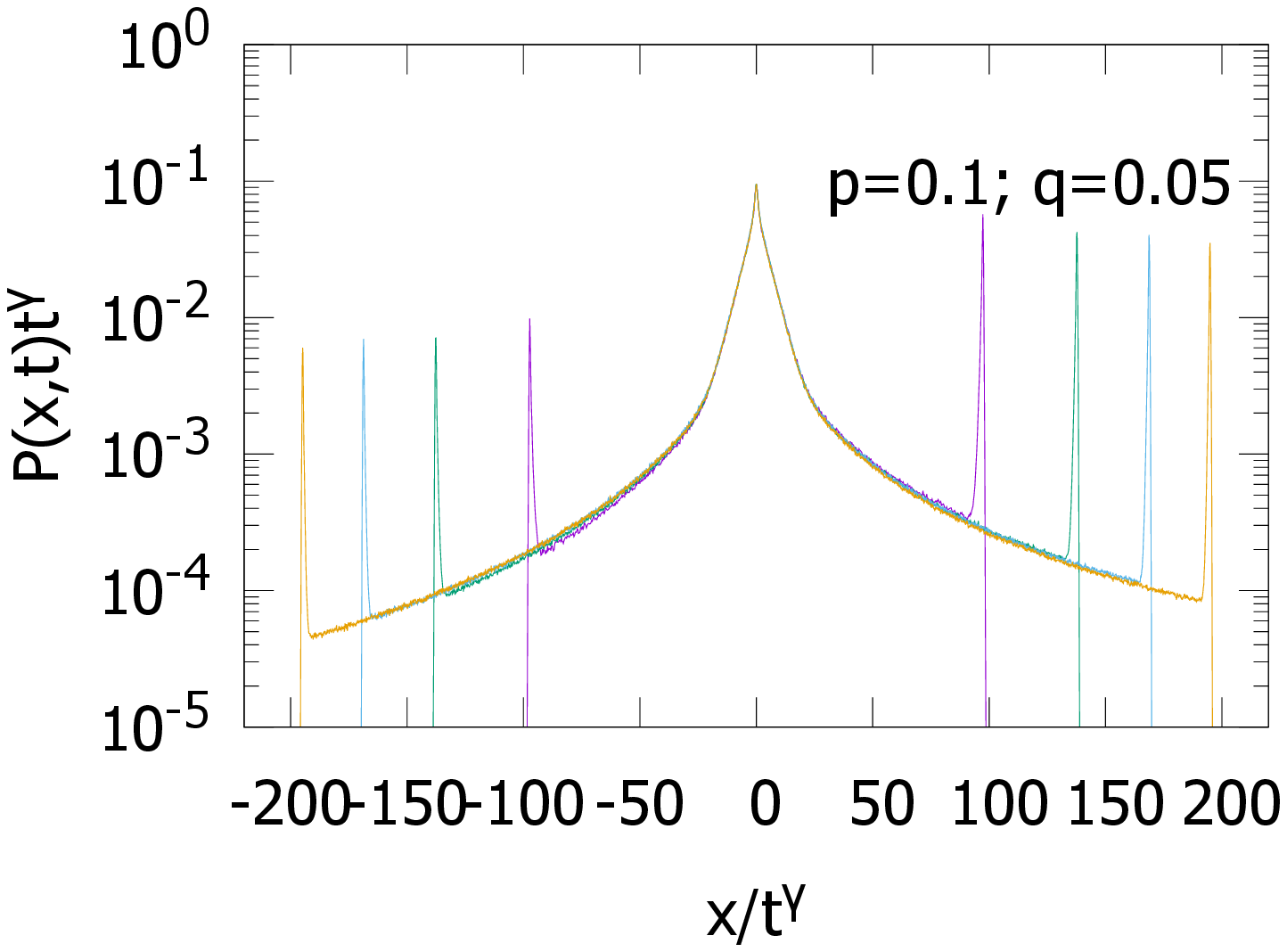}
\includegraphics[width = 0.45\linewidth]{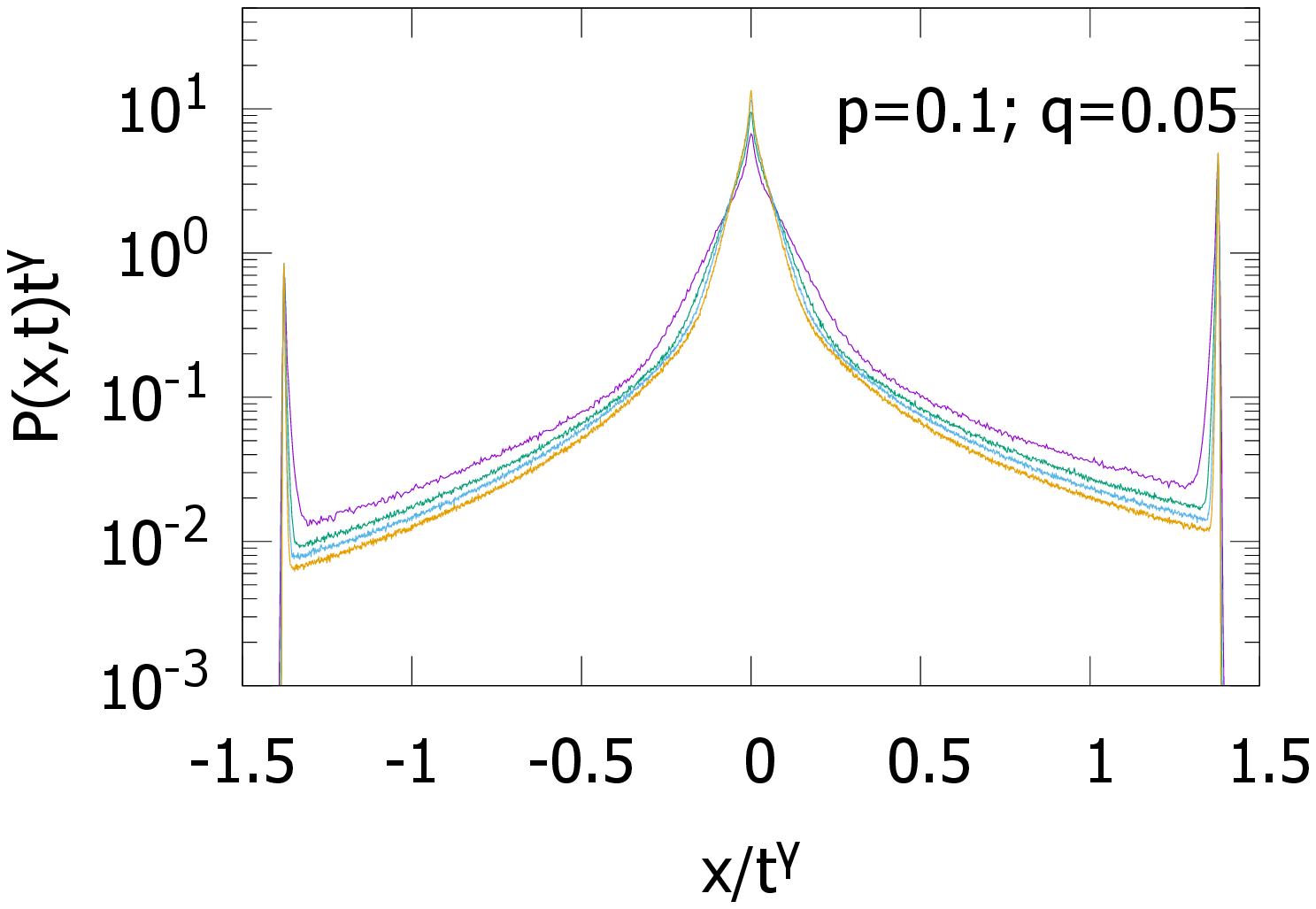}

\includegraphics[width = 0.45\linewidth]{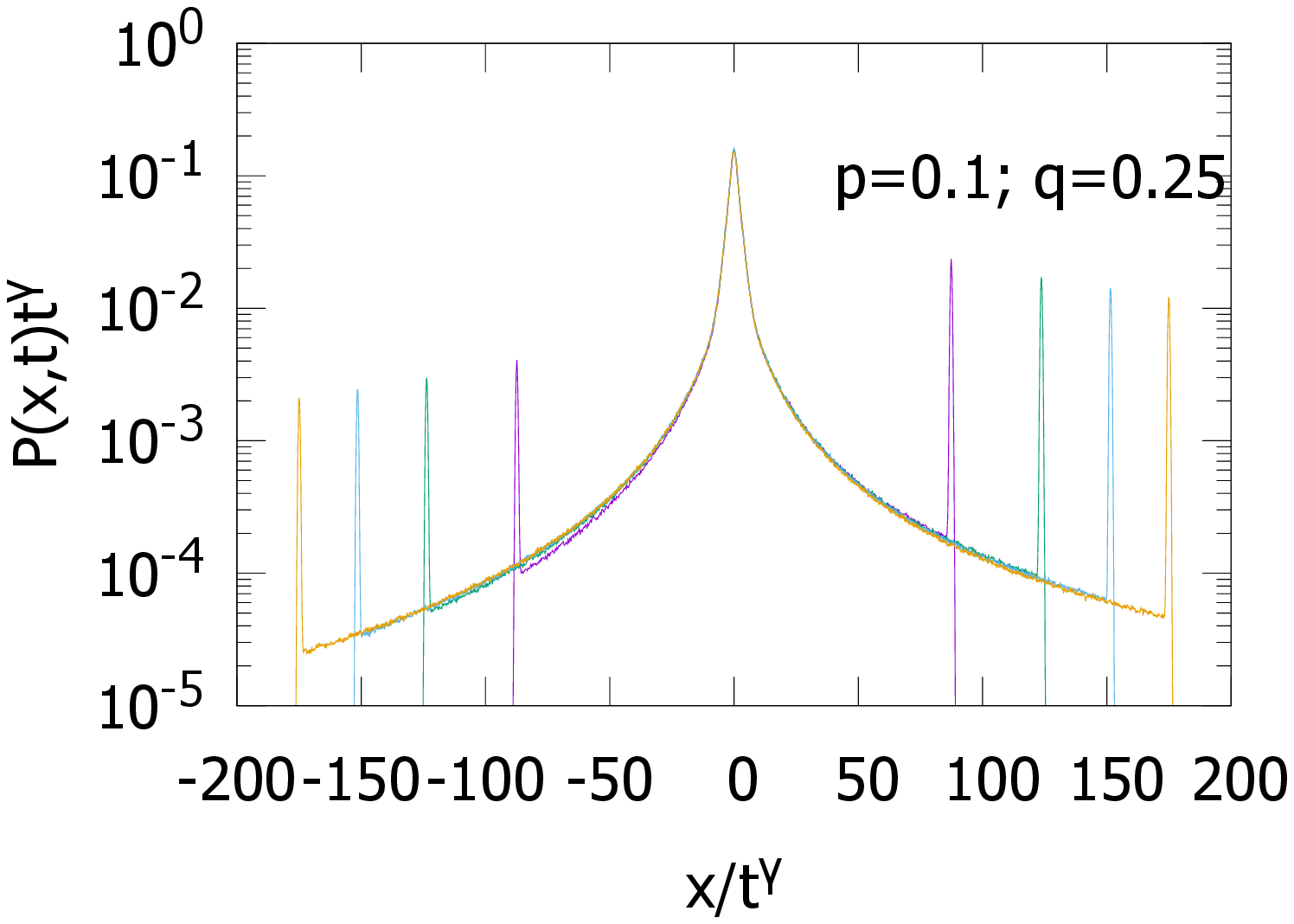}
\includegraphics[width = 0.45\linewidth]{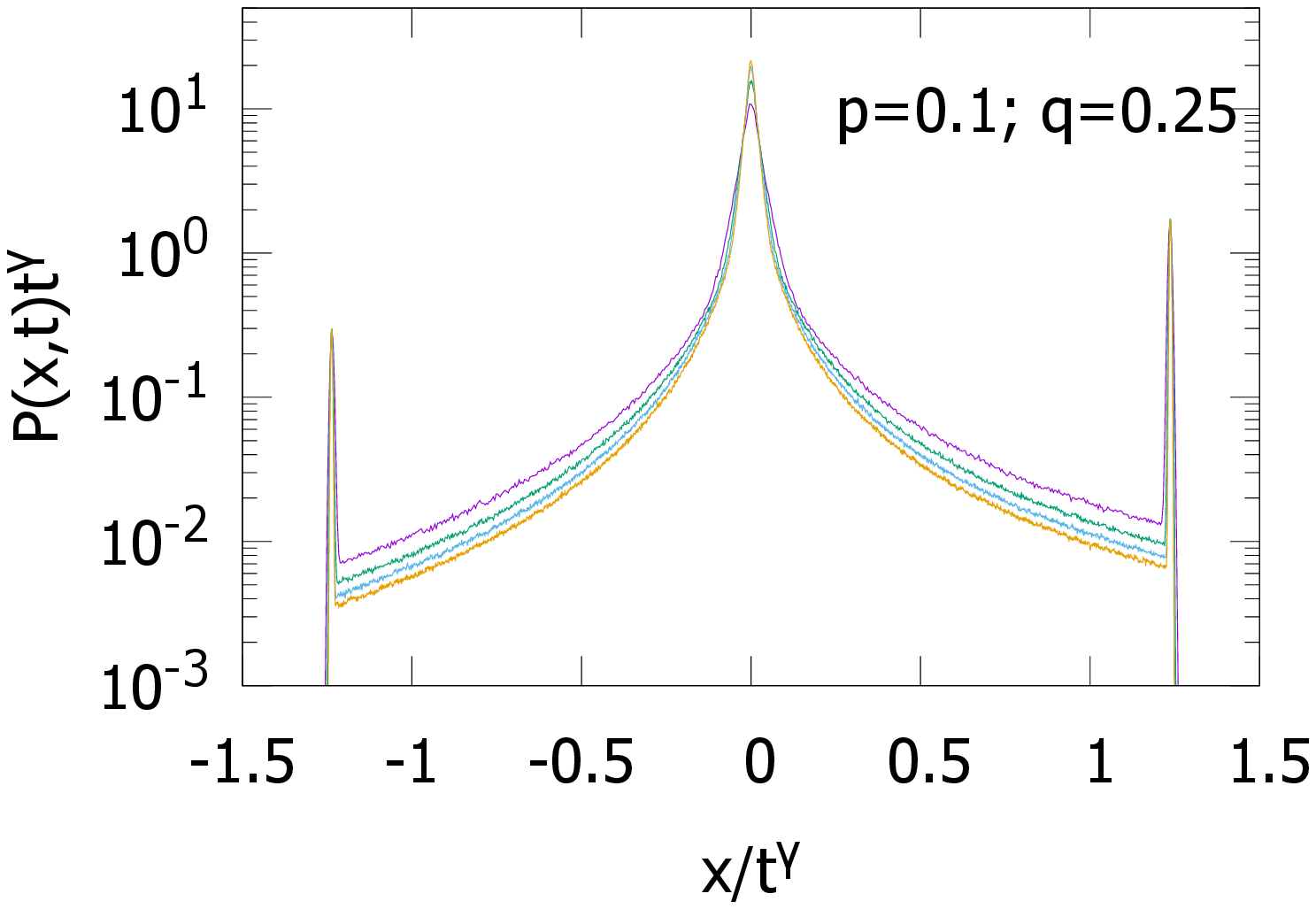}

\includegraphics[width = 0.45\linewidth]{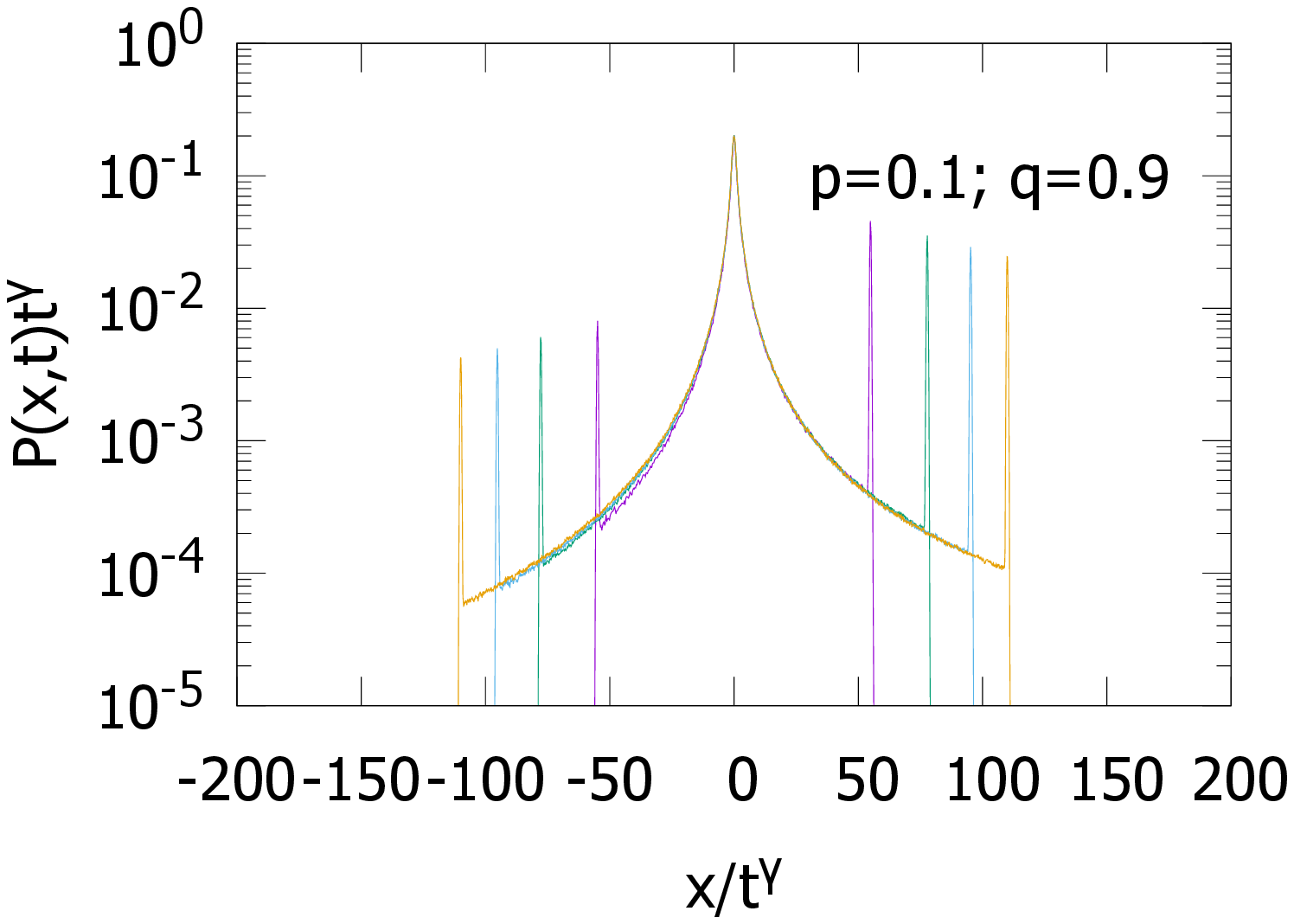}
\includegraphics[width = 0.45\linewidth]{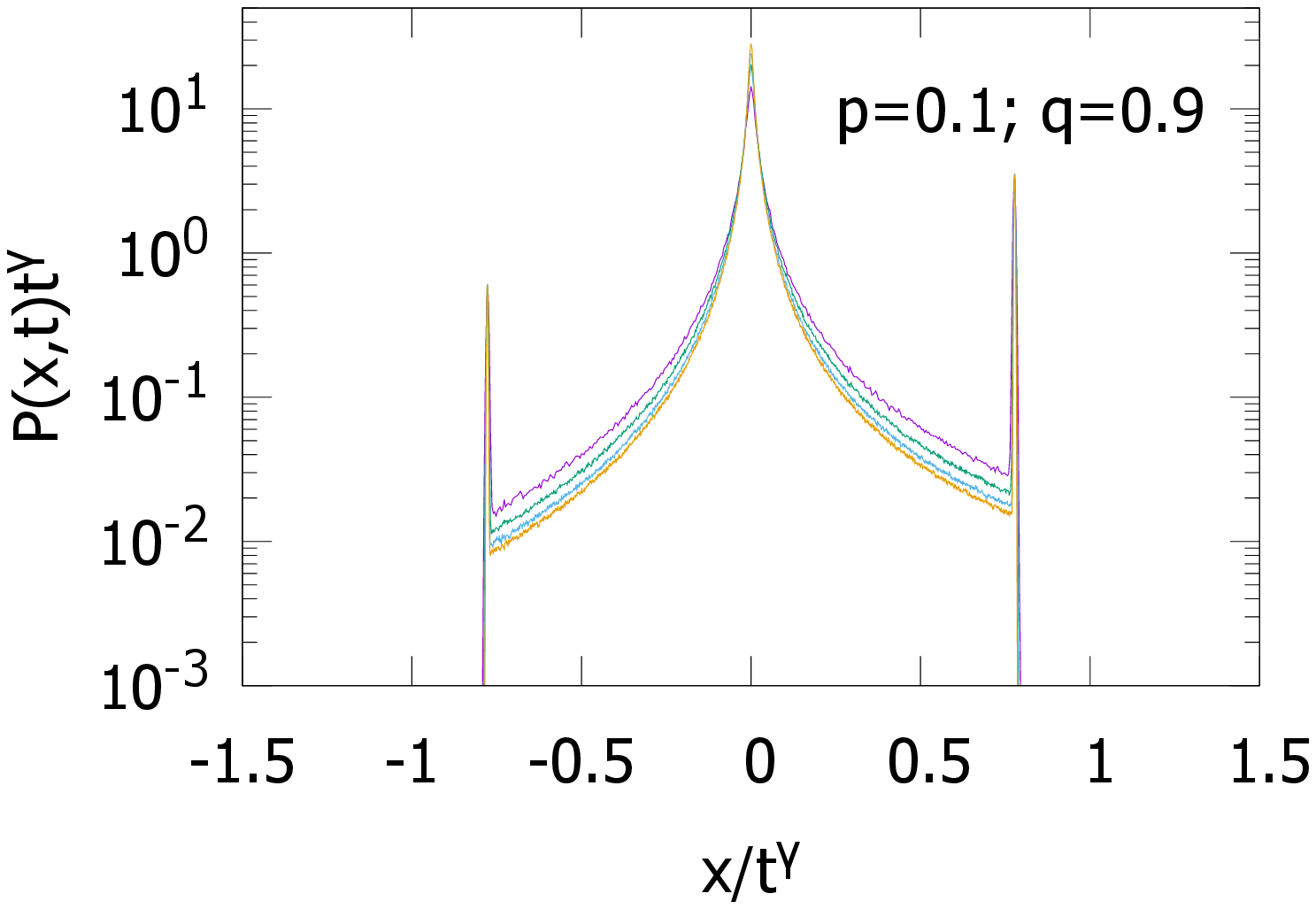}

\caption{Scheme II: Data collapse of rescaled $P(x,t)$   using $\gamma = 0.5$ (left column) and $\gamma = 1.0$ (right column). $\gamma = 1$ collapse is less sharp.}
\label{fig7:dist-scheme2}
\end{figure}

Once again we obtain the distribution $P(x,t)$ which shows a peak centered at the origin and two ballistic peaks.
As seen for Scheme I, data collapse is observed with $\gamma =  0.5$ for the central peak and $\gamma = 1.0$ for the ballistic peaks (Fig. \ref{fig7:dist-scheme2}) indicating two distinct scaling behaviors $x \propto \sqrt{t}$ and $x \propto t$ respectively.

\begin{figure}[h!] 
\includegraphics[scale = 0.5]{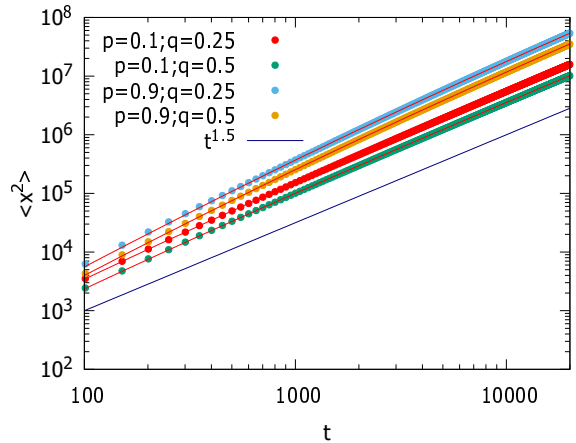}
\caption{Scheme II: The second moment for a combination of values of $p$ and $q$.  The continuous lines are best fit curves obtained using Eq. \ref{crsvr}.}
\label{fig8:mom-scheme2}
\end{figure}

Next we show the variation of the second moment in Fig. \ref{fig8:mom-scheme2} against time which shows the unique exponent $3/2$ 
asymptotically for all $ 0 < p< 1$. 
Here we have plotted the results for both small and large values of $p,q$ showing no significant  difference. 
The   exponent $\nu$  as a function of  $p$ for  $q = 0.5$ and a different value  of $q$ plotted  in Fig. \ref{fig5:nu}  shows that it is 3/2 for the entire region $0 < p< 1$, 
confirming that it is equivalent to  Scheme I when the persistent 
probability  for the latter is above $p_c$.

Lastly, we plot $S_E(t)$ against time $t$ for chosen values of $p$ and $q$ (Fig. \ref{fig6:entang}) which does not show any distinguishing feature from Scheme I.


\section{Summary and discussions}

In the present work we have reported the  results on a non-Markovian quantum walk where the  
 step lengths are binary at each time step. 
It is non-Markovian in the sense that the walker remembers the step length taken in the previous step and tends to repeat it with
probability $p$. 
Thus the choice of the step length is entirely determined  by the value of $p$.
We numerically evaluate the time evolution of the walk
and calculate the distribution $P(x,t)$ and its moments.  Scheme I, which is the case when it is strictly antipersistent with 
probability $(1-p)$,   leads to some non-intuitive results when $p$ is small.
Precisely, we find  how the  asymptotic behaviour  of the second moment changes 
as $p$ is varied. One can locate four different phases through which $\nu$ changes. The first is the point $p=0$, where the walk is periodic  and correlated over infinite time range and $\nu=2$. 
At $p=0$, there is however, a finite discontinuity in $\nu$ as $\nu =1$ for $p = 0^+$. 
For $0 < p < p_c$, $\nu$ shows a continuous increase with $p$. Here the walk 
deviates from its periodic nature and the  step lengths are  e.g., $1,2,1,2,.....1,2,2,1,2,1,......,1,1,2,1,2.....$ etc.,   such that it   
has two  ``opposite" patterns repeating  alternately. However, the long time correlation is weakened. As $p$ increases, the step lengths tend to repeat, however,  none of the 
strings, either e.g., $1,1,1...1$ or $1,2,1,2,1,2$  can have a very large length when  
 $p$ is not equal to $1$ or zero.   
As a result,  one gets approximately a  random sequence  of step lengths. In fact  at 
$p = 0.5$ it is purely random. This randomness continues to dominate unless $p$ is exactly equal to 1 so that we get $\nu = 3/2$ for $p_c < p < 1$ and $\nu = 2$ again at $p =1$.

The result for the region $0 < p < p_c$ is perhaps the most interesting where we find a $p$ dependent value of $\nu < 3/2$ that   indicates that the localisation is stronger compared to a random choice of the step lengths. 
The antipersistence effect here is able to confine the walk to a narrower region.

Let us further review the situation close to the extreme values of $p$ regarding the  
 step lengths $1$ and $2$ as up/down states of a Ising spin. 
The two sequences $1,2,1,2....$ and $2,1,2,1....$ are like antiferromagnetic patters and  are equivalent to  simply spin flipped versions of one another in this picture. 
Hence close to $ p = 0$ we have two alternating antiferromagnetic patterns and close to $p = 1$ we will have two alternating ferromagnetic patterns
separated by domain boundaries. Interestingly  for $p \to 0$ and $p \to 1$  the exponents have different values, indicating the antiferromagnetic patterns are responsible for a stronger confinement of the walk. 
For close to $p=1$, the confinement comes only from the fact that the repetition of the two ferromagnetic patterns is in no way periodic in nature.
However, why the antiferromagnetic and ferromagnetic sequences lead to different scaling behaviour and why $p_c$ is close to 1/3 remain issues 
 to be resolved.
One other result is that in general the first moment shows a scaling $\langle x\rangle \propto t^{\nu -1}$ which is not quite obvious. 

The second important result we obtain is the crossover phenomena near $p=0$ and $1$. 
Simple power law scalings for the moments are not possible here as clearly the behaviour changes in time. Of course, one can continue the numerical evolution 
for even larger number of time steps and extract the asymptotic variation. In practice, it is beyond the  computational capacity to do so. However, identification of the scaling variable and consequently obtaining a form of the  scaling function   could help in calculating the asymptotic exponent. 
We could detect the presence of  timescales which diverge at the extreme  limits and 
in the process reveal that the crossover phenomena takes place here with a 
diverging time scale.
This divergence signifies that although the exponent $\nu$ 
changes discontinuously as  
 $p \to 0$ or $p \to 1$,  
the change  can be observed only after long time scales. 
Away from the extreme limits, the crossover effect becomes less conspicuous.

The results for the persistent quantum walker shows that it is clearly different from the case of random choice of step lengths as long as antipersistence
is strong, resulting in a nonuniversal value of the exponent. This is all the more evident from the results of Scheme II in which 
the effective persistence probability corresponds to   that in Scheme I with $ p > 1/2$  and hence corresponds to the random case with  $\nu = 3/2$. 
Although the scaling exponent $\nu$ is $p$ dependent for $p < p_c$, the $P(x,t)$  data show collapse with the same type of rescaling as in the random case.  
Also, the results for the entanglement entropy are qualitatively similar to that of the latter case.

Acknowledgement:

SM is grateful for the opportunity to work in the Department of Physics, 
University of Calcutta. PS is grateful to SERB scheme number:
EMR/2016/005429.

\bibliographystyle{apsrev4-1}
\bibliography{reference}
\end{document}